\documentstyle[12pt]{article}
\title{2D shift of the centre of gravity of the light beam
carrying orbital angular momentum, which accompanies reflection
from a lossy medium}
\author{V.G.Fedoseyev\\
Institute of Physics, University of Tartu, \\
Riia 142, 51014 Tartu, Estonia}
\date{\it\today}
\begin{document}
\maketitle
\begin{abstract}
It is shown that after reflection from a lossy medium the $s$- or
$p$-polarized paraxial light beam carrying the orbital angular
momentum suffers the 2D shift of the beam's centre of gravity
relative the geometric optic axis. The mutually orthogonal
components of this shift are expressed through the real and
imaginary parts of the common complex quantity. The features of
the 2D vector, which describes the shift, are analyzed.
\end{abstract}



\section*{I. Introduction}
In the communication [1], a new nonspecular effect, which
accompanies the process of partial reflection of a paraxial light
beam at a plane sharp interface of two isotropic transparent
media, has been predicted. It was shown that, if the incident beam
carries the intrinsic orbital angular momentum (OAM), the centre
of gravity of the reflected beam (CGRB) suffers a transverse shift
(TS), i.e. the shift perpendicular to the plane of incidence.
Bekshaev and Popov [2] considered the OAM-dependent TS using the
method of transformation of space-angle intensity moments during
the interaction of the beam with a layered structure. Bliokh
compared this effect with the phenomena accompanying
transformation of the OAM in the course of propagation of a beam
in the smoothly inhomogeneous media [3].

Unlike the previously known, spin-dependent TS (see, for instance,
[4-9] and references therein), the OAM-dependent TS can take
place, when the incident beam is $s$- or $p$-polarized. This
effect appears because the electromagnetic energy inside the beam
is redistributed after reflection in such a way that the intensity
increases at one side of the incidence plane and decreases at the
other [10,11]; in the regular region, the respective corrections
are proportional to the ratio of the wavelength to the radius of
the beam.

The magnitude of the OAM-dependent TS is, if an angle of incidence
is not close to the critical angle for total reflection or (for
$p$-polarization) to the Brewster angle, rather small: it is of
the order of the wavelength or less. But, at the present, the
optical method has been developed, which permit to detect the
positions of the objects with nanometer-scale precision (see, for
instance, [12] and references therein). This method has been used
by Dasgupta and Gupta [13] in order to define the OAM-dependent
TSs of the CGRB in the wide region of the angles of incidence, the
authors of [13] have reported a good agreement between their
experimental results and the predicted values of the TSs.

Recently, Okudo and Sasada have observed the redistribution of the
electromagnetic energy inside the beam carrying the OAM after its
reflection from the transparent medium [11]. Their experiments
were carried out in a situation, where the angle of incidence was
close to the critical angle for total reflection. In this region,
the magnitude of the effect increases significantly, so the energy
redistribution becomes visible (see Figs. 3 and 4 in [11]). The
authors of [11] have performed the respective numerical
simulations, which turned out to be in a good agreement with the
experimental results. The OAM-dependent TS was not calculated in
[11], but this calculation was possible [14].

Basing on the results of [11] and [13] one can conclude that the
OAM-dependent TS of the CGRB is a quantity, which can be detected
experimentally, so, this effect can be used in order to
investigate some features of the light and the matter. Thus, this
effect is connected with the rotation motion of the
electromagnetic energy inside the incident beam [15-17]; hence, if
the OAM-dependent TS of the CGRB is detected, one can conclude
that such a motion takes place.

In this communication, we shall consider the OAM-dependent TS of
the CGRB together with another shift, which is parallel to the
plane of incidence and which corresponds to the well-known
Goos-H\"anchen shift (see, for instance, [5,8,9,18-21] and
references therein); these shifts are of the same scale, moreover,
they have a common origin. By this, we will assume that the
reflecting medium may be lossy.

\section*{II. Geometry of reflection }
We shall consider the reflection of a monochromatic light beam at
a plane interface of two semi-infinite isotropic, nondispersive,
and nonmagnetic media. The scheme of the process is shown in Fig.
1. The position of the interface is defined by equation $\hat{\bf
N}\cdot {\mbox{\boldmath $\rho$}}=0$, where ${\mbox{\boldmath
$\rho$}}$ is the 3D radius vector, and $\hat{\bf N}$ is the unit
normal to the interface directed from the first (incident) medium
to the second (reflecting) one. The dielectric constants of the
media in the upper half-space ($\hat{\bf N}\cdot {\mbox{\boldmath
$\rho$}}<0$) and the lower half-space ($\hat{\bf N}\cdot
{\mbox{\boldmath $\rho$}}>0$) will be denoted by $\epsilon_1$ and
$\epsilon_2$, respectively. The relative dielectric constant
$\epsilon =\epsilon_2/\epsilon_1$. The first medium is assumed to
be transparent, while the second medium may be lossy, i.e. we
assume that $\epsilon_1$ is positive, while $\epsilon _2$ and, as
a consequence, $\epsilon$ may be complex, so, in general case,
$\epsilon =\epsilon^{\prime}+i\epsilon ^{\prime\prime}$. Here and
further on, the prime and double prime stand for the real and
imaginary parts of a quantity.

Throughout this paper, we shall use the superscripts $i$ and $r$
for indicating the quantities characteristic of the incident and
of the reflected beams, respectively; the superscript $a$ will be
used in order to designate an arbitrary (incident or reflected)
beam, so, $a$=$i$ or $r$. We shall employ two coordinate systems
connected with the incident and the reflected beams, these systems
will be termed the $i$- and $r$-systems.

In the $i$-system, the $z\sp{(i)}$ axis is assumed to coincide
with the incident beam's axis. This axis and the unit vector
$\hat{\bf N}$ define the position of the beam's plane of incidence
in 3D space; the angle $\theta$ between the unit vectors $\hat{\bf
N}$ and $\hat{\bf z}\sp{(i)}$ is the beam's angle of incidence:
$\theta = \arccos (\hat{\bf N}\cdot \hat{\bf z}\sp{(i)})$. The
coordinate origin  $O$ is taken to be the point of intersection of
the incident beam axis with the interface. In the $r$-system, the
$z\sp{(r)}$ axis is assumed to coincide with the geometric optic
(GO) axis of the reflected beam. The latter is defined as a ray,
which is intersected by the interface at the coordinate origin,
and whose direction is characterized by the unit vector $\hat{\bf
z}\sp{(r)}$, which is related to $\hat{\bf z}\sp{(i)}$ through the
Snell's law and is written as follows: $\hat{\bf
z}\sp{(r)}=\hat{\bf z}\sp{(i)}-2(\hat{\bf N}\cdot \hat{\bf
z}\sp{(i)})\hat{\bf N}$. The beam's angle of reflection
$\theta\sp{(r)}=\arccos (\hat{\bf N}\cdot \hat{\bf
z}\sp{(r)})=\pi-\theta$.

The $y$ axis   in every system is defined to be perpendicular to
the plane of incidence, it is characterized by the unit vector
$\hat{\bf y}=\hat{\bf N} \times \hat{\bf z}\sp{(i)} / |\hat{\bf N}
\times \hat{\bf z}\sp{(i)}|$; this axis (the transverse one) is
common for both systems. The direction of the $x\sp{(a)}$ axis of
the $a$-system, which lies in the plane of incidence, is defined
by the unit vector $\hat{\bf x}\sp{(a)} =\hat{\bf y} \times
\hat{\bf z}\sp{(a)}$.

In the $a$- system, the 3D radius-vector is represented as
follows:
\begin{equation}
{\mbox{\boldmath $\rho$}}=z\sp{(a)}\hat{\bf z}\sp{(a)}+ {\bf
u}\sp{(a)}\;  ,
\end{equation}
where ${\bf u}\sp{(i)}$ and ${\bf u}\sp{(r)}$ are the 2D planar
radius vectors lying in the planes, which are perpendicular to the
axis of the incident beam and to the GO axis of the reflected
beam, respectively,
\begin{equation}
{\bf u}\sp{(a)}= x\sp{(a)}\hat{\bf x}\sp{(a)}+y\hat{\bf y}\; .
\end{equation}

\section*{III.  Incident and reflected fields}

Let us assume that the incident beam is paraxial, that it carries
the well-defined OAM ([22,23]), and that it is mainly $s$- or
$p$-polarized. Due to first assumption, the relation
\begin{equation}
\lambda /\pi b\ll 1\; ,
\end{equation}
must be fulfilled, where $\lambda$ is the wavelength of light in
the first medium, and $b$ is the beam radius.

The electrical field vector of the incident beam under
consideration can be written, in its own coordinate system, as
follows [24]:
\begin{equation}
{\bf E}_{\alpha}\sp{(i)}({\mbox{\boldmath $\rho$}})\simeq \left
[\hat{\bf e}_{\alpha}\sp{(i)}- i\frac{\lambda }{2\pi}\hat{\bf
z}\sp{(i)}\left (\hat{\bf e}_{\alpha}\sp{(i)}\cdot \frac{d}{d{\bf
u}\sp{(i)}}\right )\right ] F\sp{(i)}({\mbox{\boldmath
$\rho$}})+c.c. \; ,
\end{equation}
where the subscript $\alpha$  denotes the polarization of the
incident beam ($\alpha =s$ or $p$), while $\hat{\bf
e}_{\alpha}\sp{(i)}$ is the polarization vector: $\hat{\bf
e}_{s}\sp{(i)}\equiv\hat{\bf y}$, and $\hat{\bf
e}_{p}\sp{(i)}\equiv\hat{\bf x}\sp{(i)}$. The function
$F\sp{(i)}({\mbox{\boldmath $\rho$}})$ in the right-hand side of
Eq. (4) can be represented as a product of two terms:
\begin{equation}
F\sp{(i)}({\mbox{\boldmath $\rho$}})=f(u\sp{(i)},z\sp{(i)}) \exp
(-il\varphi\sp{(i)})\; ,
\end{equation}
where $l$ is the azimuthal index ($l=0,\pm 1,\pm 2...$), and
$\varphi\sp{(i)}$ is the azimuth obtained from the relation:
$\tan\varphi\sp{(i)}=y/x\sp{(i)}$. The term
$f(u\sp{(i)},z\sp{(i)})$ does not depend on $\varphi\sp{(i)}$, for
instance, it may be given by the Laguerre-Gaussian function or,
like in [11], by a superposition of such functions. For the
forthcoming analysis , the specification of
$f(u\sp{(i)},z\sp{(i)})$ is not necessary. The time dependence of
${\bf E}\sp{(i)}$ is suppressed.

The electrical field vector of the reflected beam ${\bf
E}\sp{(r)}({\mbox{\boldmath $\rho$}})$ is calculated in the
standard way (see, for instance, [6,10,18]). First, the 2D Fourier
transform of the function $F\sp{(i)}({\mbox{\boldmath $\rho$}})$
with respect to the coordinates $x\sp{(i)}$ and $y$ is to be
performed. Then the vector ${\bf E}\sp{(i)}({\mbox{\boldmath
$\rho$}})$ will be represented as a superposition of the plane
waves. Next, the Snell and Fresnel laws should be applied to the
particular plane waves. After that, the reverse 2D Fourier
transform should be performed.

Let us denote the angle of incidence of the particular plane wave
inside the incident beam by $\theta\sp{(ip)}$. For the actual
waves constituting a paraxial beam,
$|\theta\sp{(ip)}-\theta|\sim\lambda/b\ll 1$. So, one can expand
the particular Fresnel field reflection coefficients in the
truncated power series with respect to $\theta\sp{(ip)}-\theta$.
When the second medium is transparent, and $\epsilon^{\prime}<1$,
such an expansion can be substantiated, if $\theta$ is not too
close to the critical angle for total reflection given by
\begin{equation}
\theta\sp{(c)}=\arcsin\sqrt {\epsilon^{\prime}}\; ,
\end{equation}
i.e. if the condition
\begin{equation}
|\theta-\theta\sp{(c)}|\gg \lambda /b , \; \; \; \, if\;
\epsilon^{\prime}<1, \;\;and\;\;
\epsilon^{\prime\prime}\rightarrow-0\; ,
\end{equation}
is fulfilled. Again, in the case of $p$-polarization, the
expansion is substantiated, if $\theta$ is not too close to the
Brewster angle given by
\begin{equation}
\theta\sp{(B)}=\rm{arctan}\sqrt{\epsilon^{\prime}}\; ,
\end{equation}
i.e. if the condition
\begin{equation}
|\theta-\theta\sp{(B)}|\gg \lambda /b , \; \; \; \, if\;
\alpha\equiv p, \;\;and \;\epsilon^{\prime\prime}\rightarrow-0\; ,
\end{equation}
is fulfilled. But, when
\begin{equation}
|\epsilon^{\prime\prime}|\gg \lambda /b\; ,
\end{equation}
the series expansion is justified at any $\theta$.

Let us restrict ourselves with the near-field region (in this case
the possible angular shifts can be excluded from consideration),
i.e. let us assume that
\begin{equation}
z\sp{(r)}\ll \pi b\sp{2}/\lambda \; .
\end{equation}
If the restrictions (11), and (7), (9) or (10) are fulfilled, it
is possible to retain only the zero-order and first-order terms in
the power series expansion.

Performing the operations, which are mentioned in three by last
paragraph, one obtains the following expression for the electrical
field vector of the reflected beam written in its coordinate
system:
\begin{equation}
{\bf E}\sp{(r)}_{\alpha}({\mbox{\boldmath $\rho$}})\simeq
r_{\alpha}(\theta )\left [\hat{\bf e}_{\alpha}\sp{(r)}\left (
1+iQ_{\alpha}(\theta)\frac{\lambda}{2\pi}\frac{d}{dx\sp{(r)}}\right
)-i\frac{\lambda }{2\pi}\hat{\bf z}\sp{(r)}\left (\hat{\bf
e}_{\alpha}\sp{(r)}\cdot \frac{d}{d{\bf u}\sp{(r)}}\right ) \right
]f(u\sp{(r)},z\sp{(r)}) \exp (il\varphi\sp{(r)})+c.c. \; ,
\end{equation}
where $\hat{\bf e}_{s}\sp{(r)}\equiv\hat{\bf y}$, $\hat{\bf
e}_{p}\sp{(r)}\equiv\hat{\bf x}\sp{(r)}$, and $\tan
\varphi\sp{(r)}=y/x\sp{(r)}$. The quantity $Q_{\alpha}(\theta)$ in
the right-hand side of Eq. (12) is as follows:
\begin{equation}
Q_{\alpha}({\theta})=-\left.\frac{1}{2}\frac{d}{d\vartheta}\left
(\ln r_{\alpha}(\vartheta )\right )\right|_{\vartheta =\theta}\; ,
\end{equation}
where $r_{\alpha}(\vartheta)$ is the familiar Fresnel's field
reflection coefficient of the $\alpha$-polarized plane wave, which
is incident at an angle $\vartheta$ [25]:
\begin{equation}
r_s(\vartheta )=-\frac{\sin (\vartheta -\tau )}{\sin (\vartheta
+\tau )} \; ,\; \; \; r_p(\vartheta )=\frac{\tan (\vartheta -\tau
)}{\tan (\vartheta +\tau )}\; ,
\end{equation}
and where $\tau$ is the angle of refraction given by the relation
${\sqrt\epsilon}\sin \tau =\sin \vartheta$, in a general case
$\tau$ is complex. The dependence of the vector ${\bf
E}\sp{(r)}_{\alpha}({\mbox{\boldmath $\rho$}})$ on $\theta$ is not
explicitly written down.

Basing on Eqs. (13) and (14), the calculations of the factor
$Q_s(\theta)$ and $Q_p(\theta)$ are straightforward, the results
are as follows [10]:
\begin{equation}
Q_s(\theta)=-\frac{\sin\theta} {(\epsilon -\sin\sp 2
\theta)\sp{1/2}}\; ,
\end{equation}
and
\begin{equation}
Q_p(\theta)=\frac{Q_s(\theta)} {\epsilon\sp{-1}\sin\sp 2
\theta-\cos\sp 2 \theta}\; .
\end{equation}
Generally,  $Q_{\alpha}(\theta)$ is a complex quantity, its real
and imaginary parts are caused by the dependence on $\theta$ of
the amplitude and the phase, respectively, of the Fresnel's field
reflection coefficient $r_{\alpha}(\theta )$. The expressions (15)
and (16) are in accordance with the respective expression obtained
in [21] (see Eq. (29)).

The magnetic field vector of the reflected beam ${\bf
H}_{\alpha}\sp{(r)}({\mbox{\boldmath $\rho$}}) $ is obtained by
means of the Faraday's law.

\section*{IV. 2D shift of the CGRB}
 {\underline {\bf A. Definition and results of calculations}}.
Let us select the plane, which is perpendicular to the GO axis of
the reflected beam (the output plane in Fig. 1). The position of
the CGRB on this plane (see Fig. 2) is given by the 2D vector
\begin{equation}
{\bf U}_{\alpha}(\theta )=\frac{1}{W\sp{(r)}_{\alpha}}\int {\bf
u}\sp{(r)}w\sp{(r)}_{\alpha}({\mbox{\boldmath $\rho$}})d{\bf
u}\sp{(r)}\; ,
\end{equation}
where $w_{\alpha}\sp{(r)}({\mbox{\boldmath $\rho$}})$ is the
electromagnetic energy density inside the reflected beam
\begin{equation}
w_{\alpha}\sp{(r)}({\mbox{\boldmath $\rho$}})=\frac{1}{8\pi}\left
[\epsilon_1\left ({\bf E}\sp{(r)}_{\alpha}({\mbox{\boldmath
$\rho$}})\right ) \sp 2+\left
 ({\bf
H}\sp{(r)}_{\alpha}({\mbox{\boldmath $\rho$}})\right )\sp 2\right
] \; ,
\end{equation}
and
\begin{equation}
W_{\alpha}\sp{(r)}=\int w_{\alpha}\sp{(r)}({\mbox{\boldmath
$\rho$}})d{\bf u}\sp{(r)}\; .
\end{equation}

As the GO axis intersects the output plane at the point ${\bf
u}\sp{(r)}=0$, the vector ${\bf U}_{\alpha}(\theta )$, which is
shown in Fig. 2, describes the 2D shift of the CGRB relative to
this axis.

Let impose the following restriction on the axial coordinate of
the  output plane:
\begin{equation}
z\sp{(r)}\gg b/\sin(2\theta)\; .
\end{equation}
If the restriction (20) is fulfilled, the reflected field on the
output plane is not overlapped with the incident field. Again,
$w_{\alpha}\sp{(r)}({\bf u}\sp{(r)})$ is negligible at the line of
intersection of the output plane and the interface; in this case,
the lower limit of integration over $x\sp{(r)}$ in Eqs. (17) and
(19) can be substituted by $-\infty$. Once this substitution is
made and the expression (12) for ${\bf
E}\sp{(r)}_{\alpha}({\mbox{\boldmath $\rho$}})$ and respective
expression for ${\bf H}_{\alpha}\sp{(r)}({\mbox{\boldmath
$\rho$}})$ obtained by means of the Faraday's law are used, the
calculation of ${\bf U}_{\alpha}(\theta )\sp{(r)}$ is
straightforward. In the first-order approximation with respect to
$\lambda/b$, one obtains
\begin{equation}
{\bf U}_{\alpha}(\theta )=X_{\alpha}(\theta )\hat{\bf
x}\sp{(r)}+Y_{\alpha}(\theta )\hat{\bf y}\; ,
\end{equation}
where
\begin{equation}
X_{\alpha}(\theta )=Q^{\prime\prime}_{\alpha}(\theta
)\frac{\lambda}{\pi}\; ,
\end{equation}
and
\begin{equation}
Y_{\alpha}(\theta )=lQ_{\alpha}^{\prime}(\theta
)\frac{\lambda}{\pi}\; .
\end{equation}

{\underline {\bf B. Analysis}}. As follows from Eq. (21), the
shift of the CGRB relative to the GO axis is, in a general case,
described by the 2D vector  ${\bf U}_{\alpha}(\theta )$ on the
output plane. Its $x\sp{(r)}$-component ($X_{\alpha}(\theta )$)
does not depend on $l$, while the $y$-component
($Y_{\alpha}(\theta )$) is proportional to $l$. Apart from the
factor $l$, the functions $X_{\alpha}(\theta )$ and
$Y_{\alpha}(\theta )$ are represented by the imaginary and real
parts of the same complex quantity, it is in spite of the fact
that the mechanisms of the appearances of the orthogonal
components of the 2D shift are different [10,26]. Figs. 3(a,b)
illustrate dependence of the functions $Q^{\prime}_{\alpha}$ and
$Q^{\prime\prime}_{\alpha}$ on $\theta$ for different values of
$\epsilon^{\prime}$ and $\epsilon^{\prime\prime}$.

When the second medium is transparent, the vector ${\bf
U}_{\alpha}(\theta )$ is parallel either to the $x\sp{(r)}$ axis
(in the total-reflection regime) or to the $y$ axis (in the
partial-reflection regime). In this case, the expression (21)
reduces to the one of the previously obtained expressions, either
for the Goos-H\"anchen shift [27] or for the OAM-dependent TS of
the CGRB [1]. When the second medium is lossy, the vector ${\bf
U}_{\alpha}(\theta )$ is inclined to the coordinate axes, as is
shown in Fig. 2. By this, the direction of ${\bf
U}_{\alpha}(\theta )$ depends on the angle of incidence, and the
vector ${\bf U}_{\alpha}(\theta )$ rotates around the coordinate
origin, i.e. around the CO axis, when $\theta$ changes.

The direction of ${\bf U}_{\alpha}(\theta )$ is defined by its
azimuth $\Gamma_{\alpha}(\theta )$, which is given by the
equation: $\tan\Gamma_{\alpha}(\theta )=Y_{\alpha}(\theta
)/X_{\alpha}(\theta )$. Substituting in its right-hand side Eqs.
(22) and (23) one obtains the relation between the azimuth
$\Gamma_{\alpha}(\theta )$ and the argument of the complex
quantity $Q_{\alpha}(\theta)$, which is defined by $\gamma
_{\alpha}(\theta )=(\ln Q_{\alpha}(\theta))^{\prime\prime})$; this
relation looks as follows :
\begin{equation}
\tan\Gamma_{\alpha}(\theta )=l/\tan\gamma_{\alpha}(\theta )\; .
\end{equation}
In particular, if $l=\pm 1$,
\begin{equation}
\Gamma_{\alpha}(\theta )=l(\frac{\pi}{2}-\gamma_{\alpha}(\theta
))\; .
\end{equation}
As for the arguments $\gamma_{s}(\theta )$ and $\gamma_{p}(\theta
)$, they are obtained from the expressions (15) and (16):
\begin{equation}
\gamma_{s}(\theta )=\frac{1}{2}\rm{arccotan}\left
(\frac{\epsilon^\prime- \sin\sp 2
\theta}{|\epsilon^{\prime\prime}|}\right )-\pi \; ,
\end{equation}
and
\begin{equation}
\gamma_{p}(\theta )=\gamma_{s}(\theta )-\rm{arccotan}\left
(\frac{\epsilon^\prime- |\epsilon|\sp 2\cot\sp
2\theta}{|\epsilon^{\prime\prime}|}\right )\; .
\end{equation}
Basing on Eqs. (24)-(27) one can make the following conclusions
about the orientation of the vectors ${\bf U}_s(\theta )$ and
${\bf U}_p(\theta )$. As follows from Eq. (26), $-\pi
/2>\gamma_{s}(\theta )>-\pi$ at any $\theta$, i.e. $Q_s(\theta )$
lies in the third quadrant of the complex plane; as a consequence,
the vector ${\bf U}_s(\theta )$ (see Eq. (24))lies in the third
quadrant of the output plane ($X_s(\theta )<0, Y_s(\theta )<0$),
if $l>0$, and in the second quadrant($X_s(\theta )<0, Y_s(\theta
)> 0$), if $l<0$. The positions of $Q_p(\theta)$ on the complex
plane and of the vector ${\bf U}_p(\theta )$ on the output plane,
unlike of $Q_s(\theta)$ and ${\bf U}_s(\theta )$, are not
contained within one quadrant, when $\theta$ changes from 0 to
$\pi /2$. At small $\theta$ the direction of the vector ${\bf
U}_p(\theta )$ is approximately opposite with respect to the
direction of the vector ${\bf U}_s(\theta )$. As $\theta$
increases, the sign of $Y_p(\theta )$ changes once independent of
the value of $\epsilon$. The sign of $X_p(\theta )$ changes, if
the condition
\begin{equation}
2\epsilon^\prime>|\epsilon|\sp{2}  \;
\end{equation}
is fulfilled. The numerical analysis of dependence of $X$ on
$\theta$ has been recently performed by

The angles, at which these changes take place, are obtained from
the following equation:
\begin{equation}
|\epsilon|\sp 2\cot\sp 4\theta+2
(1+\epsilon^{\prime}|\epsilon|\sp{-2})\sin\sp{2}\theta-3=0\; .
\end{equation}
If $\epsilon^{\prime\prime}\rightarrow-0$ and
$\epsilon^{\prime}>0$, the second term in the right-hand side of
Eq. (27) have the jump discontinuity at the Brewster angle
$\theta\sp{(B)}$, where $\gamma_{p}$ changes abruptly by $\pi$. If
in addition $\epsilon^{\prime}<1$, the ($\pi /2$)-jump of
$\gamma_{s}$ and $\gamma_{p}$ takes place at the critical angle
for total reflection $\theta\sp{(c)}$. Figs. 4(a,b) illustrate
dependence of the functions $\Gamma_p$ and $\Gamma_s$ on $\theta$
for different values of $\epsilon$ and $l$.

The scale of the 2D shift is $\lambda/\pi$, but, if
$|\epsilon^{\prime\prime}|\ll 1$, its magnitude increases
significantly in the vicinity of the angle $\theta\sp{(c)}$ and,
for $p$-polarization,  in the vicinity of the angle
$\theta\sp{(B)}$, these angles are given by Eqs. (6) and (8),
respectively. Fig. 5 illustrates the behavior of the vector ${\bf
U}_{s}(\theta )$ in the former domain, the behavior of the vector
${\bf U}_{p}(\theta )$ in this domain is similar (see further Eq.
(30)). It is seen that, when $l\ne 0$ , the effect of rotation of
the vector ${\bf U}_{\alpha}(\theta )$ around the GO axis as well
as change of its length in the process of the change of $\theta$
is pronounced. Again, if $|\epsilon^{\prime\prime}|\ll 1$, and
$|\theta-\theta\sp{(c)}|\ll 1$, the position of the vector ${\bf
U}_{\alpha}(\theta )$ is sensitive to small changes of the values
of $\epsilon^{\prime}$ and $\epsilon^{\prime\prime}$ (see Figs. 6
and 7). If the above-mentioned conditions are fulfilled, the
approximate expressions of the functions
$Q_{\alpha}^{\prime}(\theta )$ and
$Q^{\prime\prime}_{\alpha}(\theta )$ look as follows:
\begin{equation}
Q_{s}^{\prime ,\prime\prime}(\theta )=
\epsilon^{\prime}Q_{p}^{\prime ,\prime\prime}(\theta )=-\left
(\frac{\epsilon^{\prime}}{2|\epsilon^{\prime\prime}|}\right
)\sp{1/2}\left (\frac{(\Delta \sp 2 +1) \sp{1/2}\mp\Delta
}{\Delta\sp 2 +1}\right )\sp{1/2}\; ,
\end{equation}
where  $\Delta =(\theta-\theta\sp{(c)})\sin
(2\theta\sp{(c)})/|\epsilon^{\prime\prime}|$, in (30) and further
on the signs minus and plus stand for $Q_{\alpha}^{\prime}(\theta
)$ and $Q^{\prime\prime}_{\alpha}(\theta )$, respectively. Notice
that, relative to the point $\theta=\theta\sp{(c)}$, i.e. to the
point $\Delta =0$, the functions $Q_{\alpha}^{\prime}(\theta )$
and $Q^{\prime\prime}_{\alpha}(\theta )$ given by Eq. (30) are
mutually symmetric. The magnitudes of these functions achieve the
maximums at the symmetrical points $\Delta_{max}^{(\prime )}$ and
$\Delta_{max}^{(\prime\prime )}$, respectively, given by
$\Delta_{max}^{(\prime\prime )}=-\Delta_{max}^{(\prime
)}=1/\sqrt{3}$. By this, $Q_{s}^{\prime}(\Delta_{max}^{(\prime
)})=Q_{s}^{\prime\prime}(\Delta_{max}^{(\prime\prime)})=-(3/4)\sp{3/4}
(\epsilon^{\prime}/|\epsilon^{\prime\prime}|)\sp{1/2}$.

The behavior of the vector ${\bf U}_{\alpha}(\theta )$ in the
vicinity of $\theta\sp{(B)}$ is shown in Fig. 8. Like in the
previous case, the effect of rotation of this vector around the GO
axis with the change of $\theta$ is pronounced in this domain;
again, the position of ${\bf U}_{p}(\theta )$ is sensitive to
small changes of the values of $\epsilon^{\prime}$ and
$\epsilon^{\prime\prime}$ (these effects are not demonstrated).
When $|\epsilon^{\prime\prime}|\ll1$, and $|\theta
-\theta\sp{(B)}|\ll 1$,
\begin{equation}
Q_p(\theta)\simeq-\frac{1}{2}\left (\theta-\theta\sp{(B)}+i\frac
{|\epsilon^{\prime\prime}|}{2(1+\epsilon^{\prime})
\sqrt{\epsilon^{\prime}}}\right )\sp{-1}\;,
\end{equation}
while $|Q_s(\theta)|$ is small in comparison with $|Q_p(\theta)|$,
namely, $Q_{s}^{\prime}(\theta )\simeq
-1/\sqrt{\epsilon^{\prime}}$, and $|Q_{s}^{\prime\prime}(\theta
)|\ll |Q_{s}^{\prime}(\theta )|$.

The numerical analysis of dependence of Goos-H\"anchen shift, i.e.
of the quantity $X$, on $\theta$ has been recently performed by
Lai {\it et al.} The authors of [21] have paid special attention
to the behavior of this shift in the vicinity of the Brewster dip.
Eq. (31) shows that, if $|\epsilon^{\prime\prime}|\ll1$, the
function $X_p(\theta )$ is approximated in this region by the
Lorentz curve.

{\underline {\bf C. Relative p-s shift}}. Detection of the shift
of the CGRB of the $s$- or $p$-polarized beam requires a careful
determination of the position of the GO axis. The latter procedure
can meet the difficulties. One can escape these difficulties by
operating with the difference of the shifts of two beams having
the orthogonal polarizations, such a scheme has been realized in
the experiments by Dasgupta and Gupta [13]. In [13], the initial
beam was randomly polarized, and the $s$- or $p$-polarization were
selected by means of rotation of a linear polarizer. The other
elements of the set-up were kept motionless.

>From Eqs. (17) and (19), the difference between the centres of
gravity of the respective reflected beams
\begin{equation}
{\bf U}_{ps}(\theta )\equiv{\bf U}_{p}(\theta )-{\bf U}_s(\theta
)=\frac{1}{W\sp{(r)}_sW\sp{(r)}_p}\int {\bf
u}\sp{(r)}G\sp{(r)}({\bf u}\sp{(r)})d{\bf u}\sp{(r)} ,
\end{equation}
where $G\sp{(r)}({\bf u}\sp{(r)})$ is the cross-correlation
function,
\begin{equation}
G\sp{(r)}({\bf u}\sp{(r)})=\int w\sp{(r)}_p({\bf u}\sp{(r)}+{\bf
v}\sp{(r)})w\sp{(r)}_s({\bf v}\sp{(r)})d{\bf v}\sp{(r)}\; ,
\end{equation}
and where ${\bf v}\sp{(r)}$ is, like ${\bf u}\sp{(r)}$, the 2D
radius vector lying in the plane perpendicular to $\hat{\bf
e}_z\sp{(r)}$. The vector ${\bf U}_{ps}(\theta )$, unlike the
vectors ${\bf U}_s(\theta )$ and ${\bf U}_{p}(\theta )$, does not
depend on the position of the GO axis.

Let us consider the case, when both incident beams have the common
axes and, as a consequence, both reflected beams have the common
GO axes. Such a situation takes place, for instance, if, within
the scheme of the experiment [13], the rotation of the polarizer
do not affect the position of the axis of the incident beam. In
this case

\begin{equation}
{\bf U}\sp{(r)}_{ps}(\theta )=\left
(Q^{\prime\prime}_{ps}(\theta)\hat{\bf
x}\sp{(r)}+lQ^{\prime}_{ps}(\theta)\hat{\bf y}\right )\frac
{\lambda}{\pi}\; ,
\end{equation}
where
\begin{equation}
Q_{ps}(\theta)=Q_{p}(\theta)-Q_{s}(\theta)\; .
\end{equation}

Generalization of the expression (34) for the case, when the axes
of the mutually orthogonally polarized incident beams are
parallel, is straightforward.

\section*{V. Conclusions}

When the $s$- or $p$-polarized paraxial light beam, which carries
the OAM, is reflected from the lossy medium, two mutually
orthogonal shifts of the CGRB to scale $\lambda$ appear; so, in
the near-field region, the total shift is described by the 2D
vector ${\bf U}_s(\theta )$ or ${\bf U}_p(\theta )$ on the output
plane. The vector ${\bf U}_{\alpha}(\theta )$, with $\alpha=s$ or
$p$, is given by Eq. (21) and its orthogonal components are given
by Eqs. (22) and (23). The second medium is transparent, these
expressions are correct, if conditions (7) and (9) are fulfilled.
The second medium is lossy, and condition (10) for
$|\epsilon^{\prime\prime}|$ is fulfilled, the above-mentioned
expressions are correct at any $\theta$.

The vectors ${\bf U}_s(\theta )$ and ${\bf U}_p(\theta )$ are
defined relative the GO axis of the reflected beam; the difference
between them (the vector ${\bf U}_{ps}(\theta )\equiv{\bf
U}_{p}(\theta )-{\bf U}_s(\theta )$) does not depend on the
position of the GO axis, it is expressed through the
cross-correlation function of two energy distributions across the
output plane (Eqs. (32) and (33)).

The component of the vector ${\bf U}_{\alpha}(\theta )$, which is
parallel to the plane of incidence ($X_{\alpha}(\theta )$)
corresponds to the Goos-H\"anchen shift, it does not depend on the
azimuthal index $l$. The transverse component $Y_{\alpha}(\theta
)$, which represents the OAM-dependent TS of the CGRB, is
proportional to $l$. Apart of the factor $l$, the components
$X_{\alpha}(\theta )$ and $Y_{\alpha}(\theta )$ are given by the
imaginary and real parts of the common quantity.

Redistribution of the electromagnetic energy inside the reflected
beam, which entails the OAM-dependent TS of the CGRB, is connected
with the rotation energy motion inside the incident beam [15-17];
hence, if a deflection of the vector ${\bf U}_{\alpha}(\theta )$
from the $x\sp{(r)}$ axis is detected, one can conclude that such
a motion takes place.

In the case of $s$-polarization the signs of both components of 2D
shift do not depend on $\theta$: the vector ${\bf U}_s(\theta )$
lies in the third quadrant of the output plane, if $l>0$, and in
the second quadrant, if $l<0$. On the contrary, the positions of
the vector ${\bf U}_p(\theta )$ on the output plane is not
contained within one quadrant when $\theta$ changes from 0 up to
$\pi /2 $. The azimuths of  the vectors ${\bf U}_s(\theta )$ and
${\bf U}_p(\theta )$ are given by Eqs. (24)-(27). At small
$\theta$ the directions of these vectors are approximately
opposite . As $\theta$ increases, the sign of $Y_p(\theta )$
changes one time independent of the value of $\epsilon$, while the
sign of $X_p(\theta )$ changes, if the condition (28) is
fulfilled. Generally, when $\theta$ changes, the rotation of the
vector ${\bf U}_{\alpha}(\theta )$ around the GO axis as well as
the change of its length takes place; these effects are most
pronounced if $|\epsilon^{\prime\prime}|\ll 1$ and $\theta$ is
close to $\theta\sp{(c)}$ or, for $\alpha =p$, to
$\theta\sp{(B)}$. In these domains, the quantity
$Q_{\alpha}(\theta)$ is approximated by expression (30) or (31).
In the vicinity of $\theta\sp{(c)}$, the dependence of the vector
${\bf U}_{s}(\theta )$ on $\theta$ is shown in Fig. 5, the
behaviors of the vectors ${\bf U}_{p}(\theta )$ and ${\bf
U}_{ps}(\theta )$ are, due to the left relation in (30), similar
to behavior of ${\bf U}_{s}(\theta )$. In the vicinity of
$\theta\sp{(B)}$, the dependence of the vector ${\bf U}_{p}(\theta
)$ on $\theta$ is shown in Fig. 8.

In the above-mentioned domains, the position of the vector ${\bf
U}_{\alpha}(\theta )$ is sensitive to small changes of the values
of $\epsilon^{\prime}$ and $\epsilon^{\prime\prime}$ (see Figs. 6
and 7). This may be used as a tool for the investigation of the
small changes of the dielectric constants of the media. Again, in
view of this fact, one can await that, if the reflecting medium is
taken to be nonlinear, the dependence of the position of the
vector ${\bf U}_{\alpha}(\theta )$ on the light intensity should
be pronounced in the above-mentioned domains.

\section{Acknowledgment.}

The author thanks H.Sasada for valuable correspondence. This work
was supported by Estonian Science Foundation (grant No. 6534)

\newpage

FIGURE CAPTIONS.

Fig. 1. Geometry of reflection.

Fig. 2. Position of the vector ${\bf U}$ on the output plane, the
azimuth $\Gamma$ is shown, the subscripts $\alpha$ are suppressed
in this figure. The numerals 1, 2, 3, and 4 denote the quadrant'
numbers.

Fig. 3(a). $Q^{\prime}_{\alpha}$ (thin curves) and
$Q^{\prime\prime}_{\alpha}$ (thick curves) versus $\theta$.
$\epsilon^{\prime}=2$; solid lines correspond to
$\epsilon^{\prime\prime}=-0.001$, and dashed lines to
$\epsilon^{\prime\prime}=-0.5$.

Fig. 3(b). Same as in Fig. 3(a) but with $\epsilon^{\prime}=0.5$

Fig. 4(a). $\Gamma_{\alpha}$ versus $\theta$.
$\epsilon^{\prime}=2$. Solid curves for $l=1$,
$\epsilon^{\prime\prime}=-0.001$,  dashed curves for $l=1$,
$\epsilon^{\prime\prime}=-0.5$, and dotted curves for $l=5$,
$\epsilon^{\prime\prime}=-0.5$.

Fig. 4(b). Same as in Fig. 4(a) but with $\epsilon^{\prime}=0.5$

Fig. 5. Dependence of the components $X_{s}$ and $Y_{s}$ on
$\theta$ in the vicinity of $\theta\sp{(c)}$.
$\epsilon^{\prime}=0.5$ ($\theta\sp{(c)}=45^{o}$), and
$\epsilon^{\prime\prime}=-0.001$. The thin, middle, and  thick
curves correspond to $l=0, 1$, and $2$, , respectively.

Fig. 6. Dependence of the components $X_{s}$ and $Y_{s}$ on
$\epsilon^{\prime}$ in the vicinity of
$\epsilon^{\prime}=\sin^{2}\theta$. $l=1$, $\theta =45^{o}$. The
thick and thin curves correspond to
$\epsilon^{\prime\prime}=-0.001$ and
$\epsilon^{\prime\prime}=-0.01$, respectively.

Fig. 7. Dependence of the components $X_{s}$ and $Y_{s}$ on
$\epsilon^{\prime\prime}$ in the situation, when $\theta$ is close
to $\theta\sp{(c)}$. $l=1$, $\epsilon^{\prime}=0.5$. The thin and
thick curves correspond to $\theta
=44.9^{o}=\theta\sp{(c)}-0.1^{o}$ and  $\theta
=45.1^{o}=\theta\sp{(c)}+0.1^{o}$, respectively.

Fig. 8. Dependence of the components $X_{p}$ and $Y_{p}$ on
$\theta$ in the vicinity of $\theta\sp{(B)}$.
$\epsilon^{\prime}=2.0$ ($\theta\sp{(B)}=54.7^{o}$), and
$\epsilon^{\prime\prime}=-0.001$. The thin and  thick curves
correspond to $l=1$, and $2$, respectively.

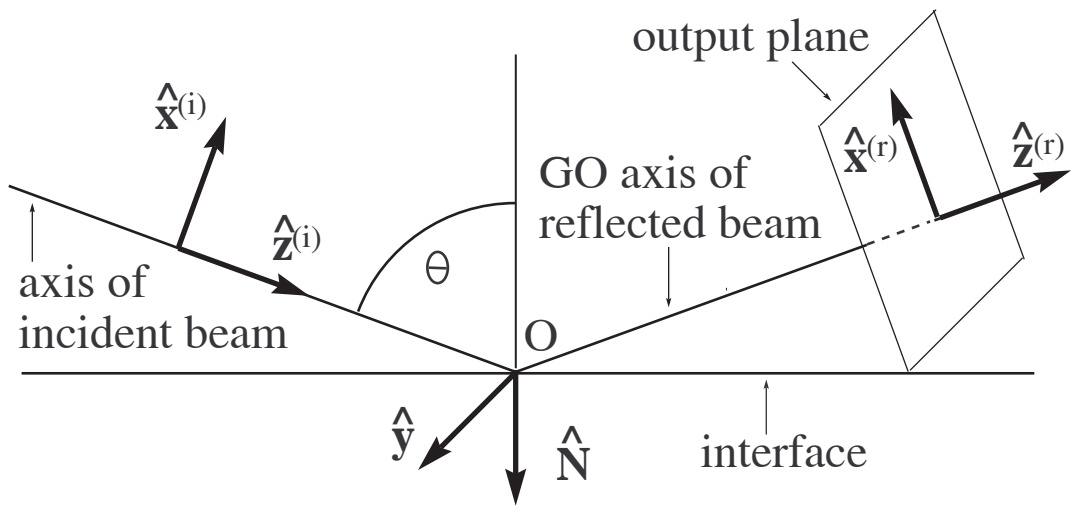
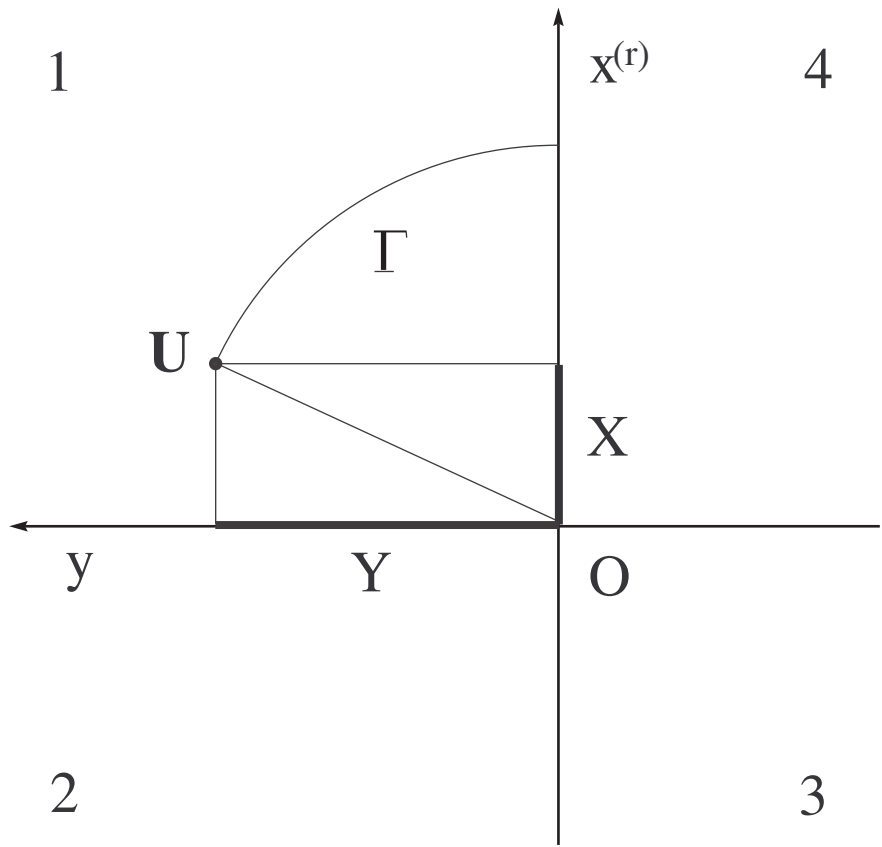
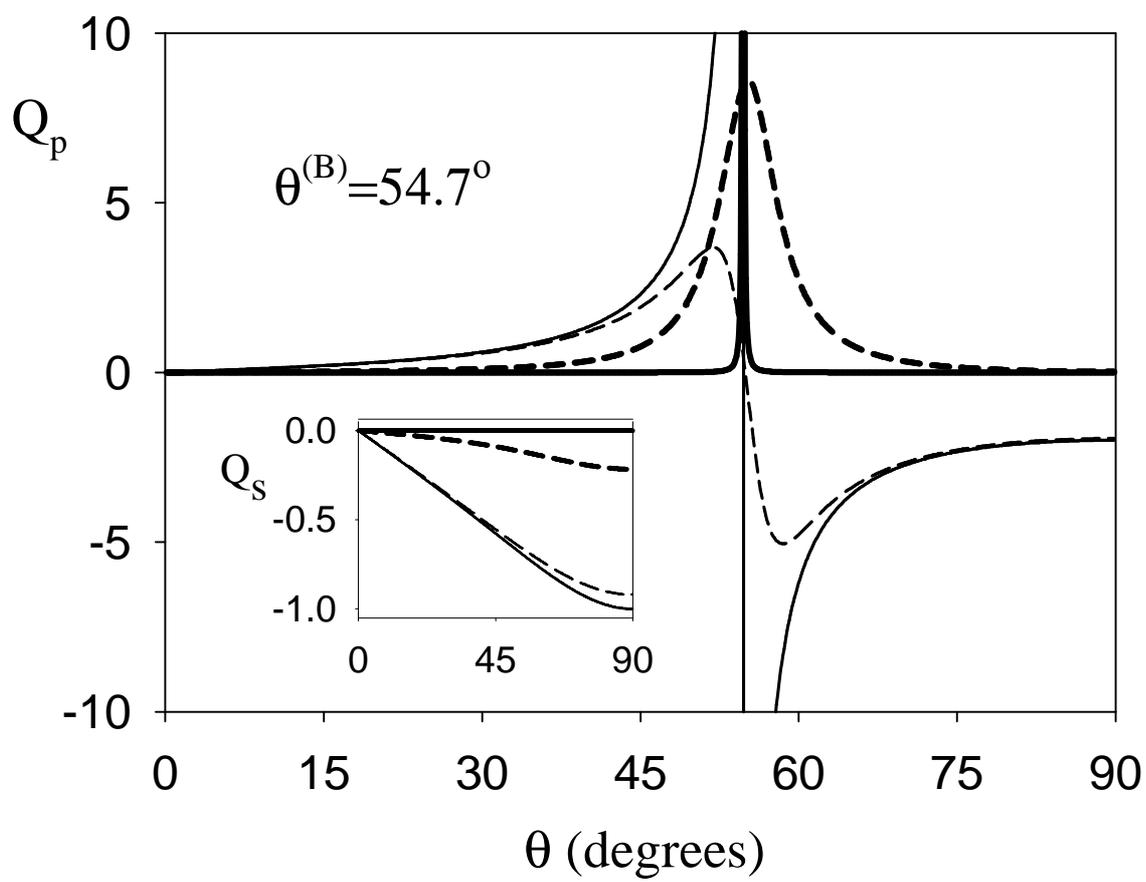
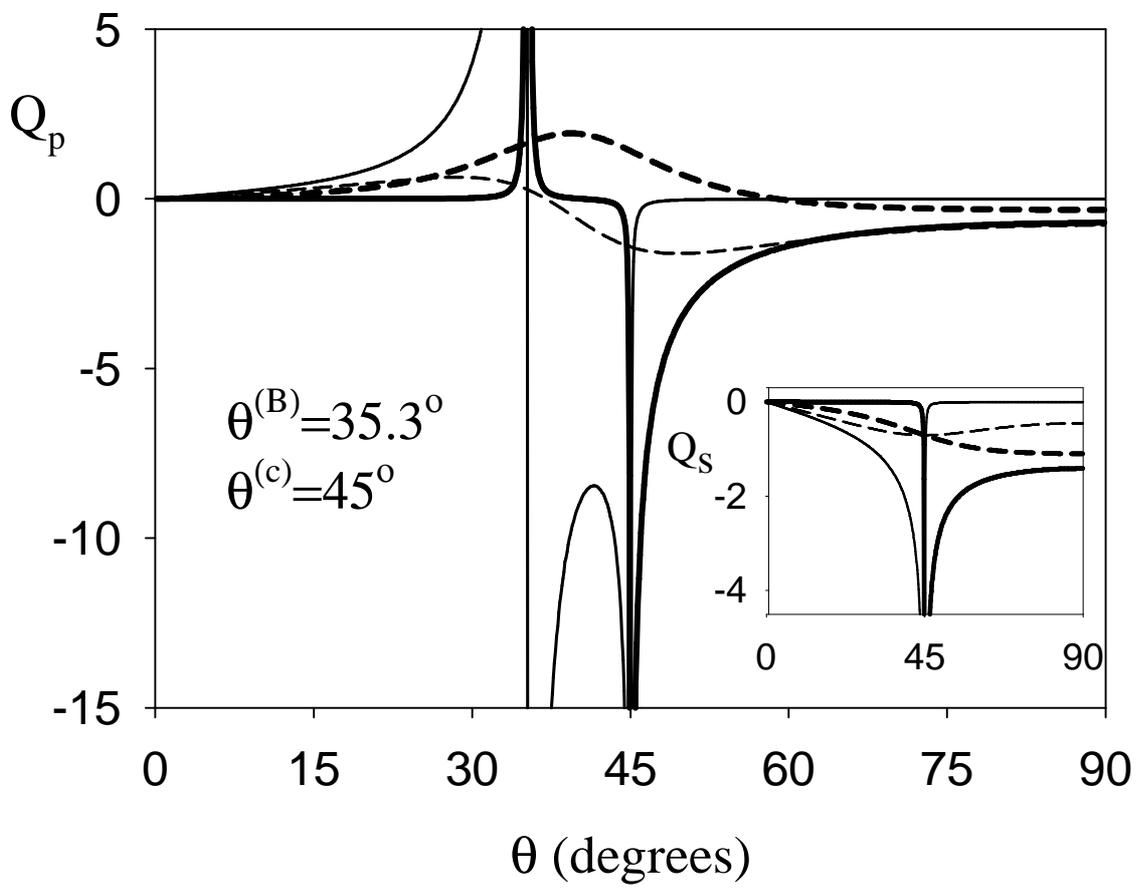
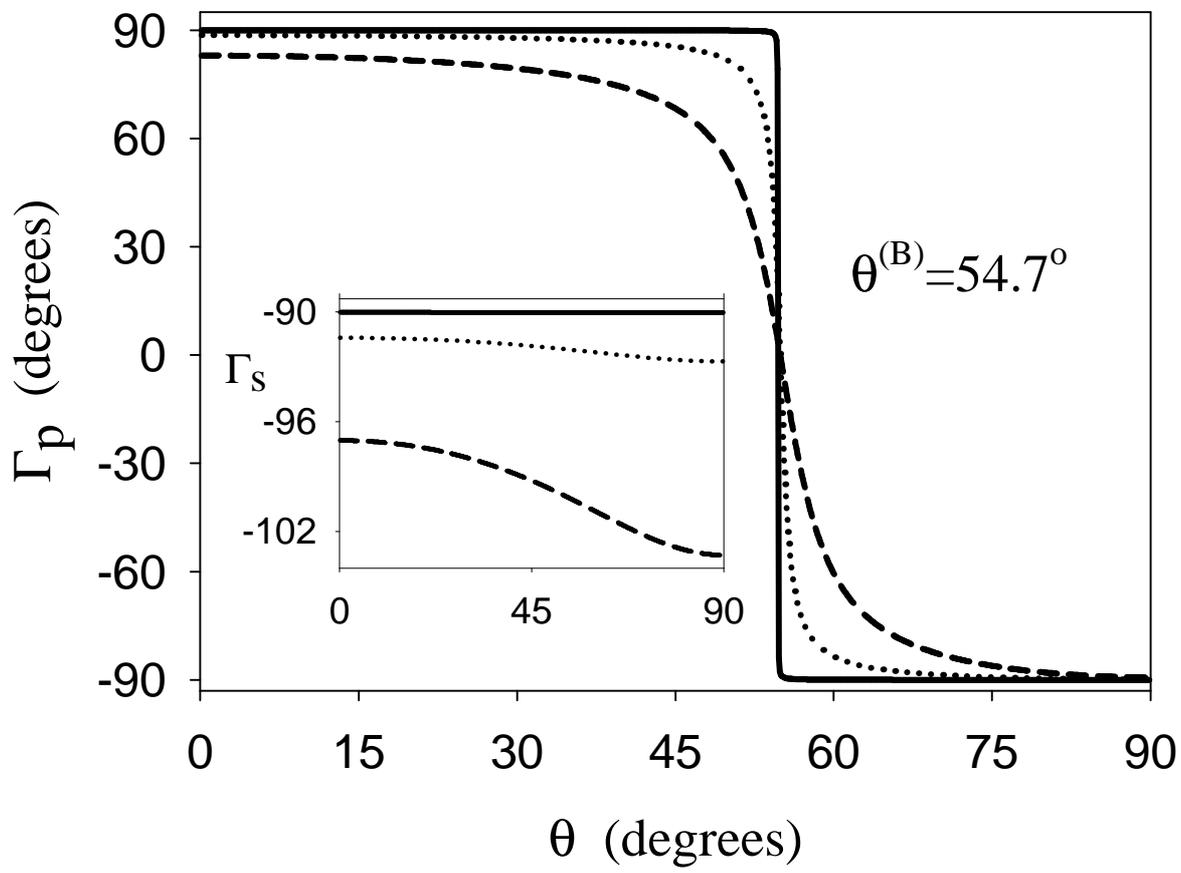
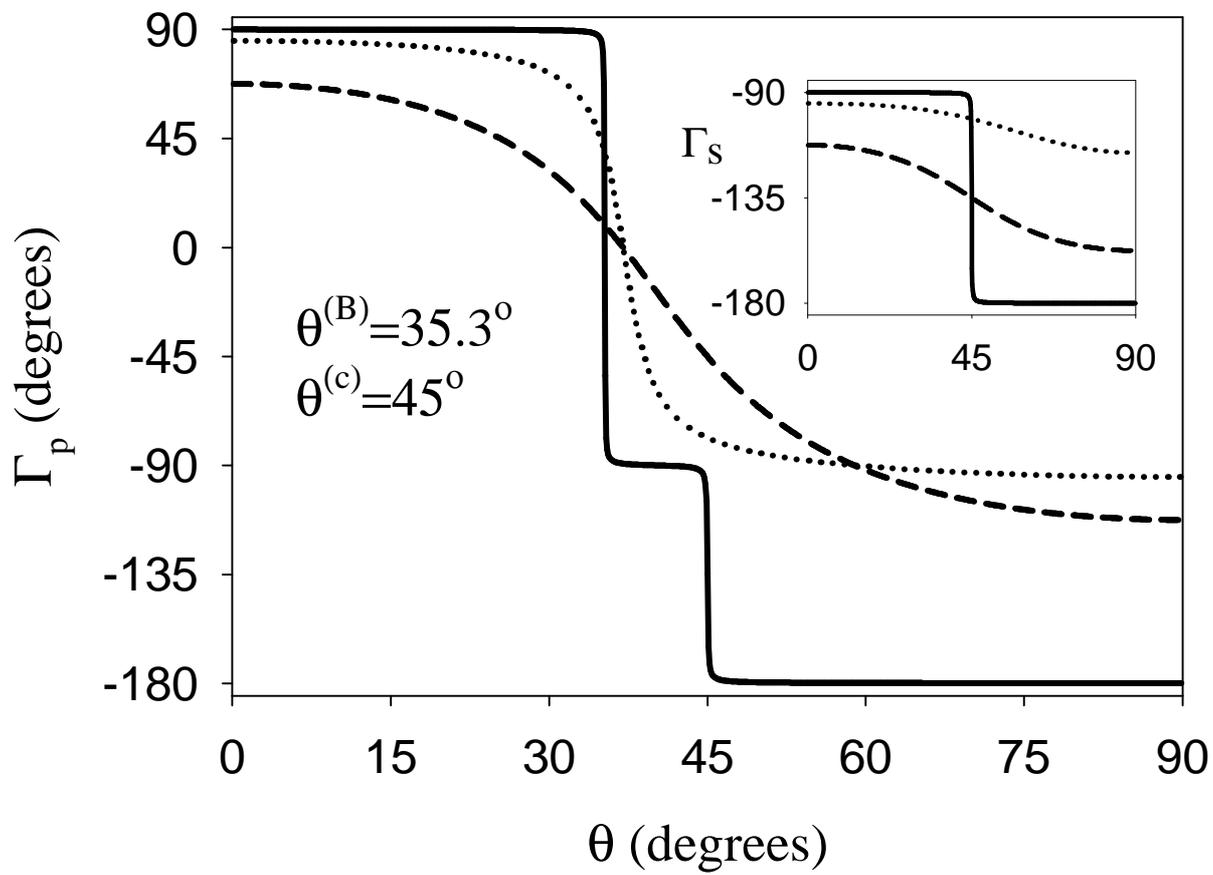
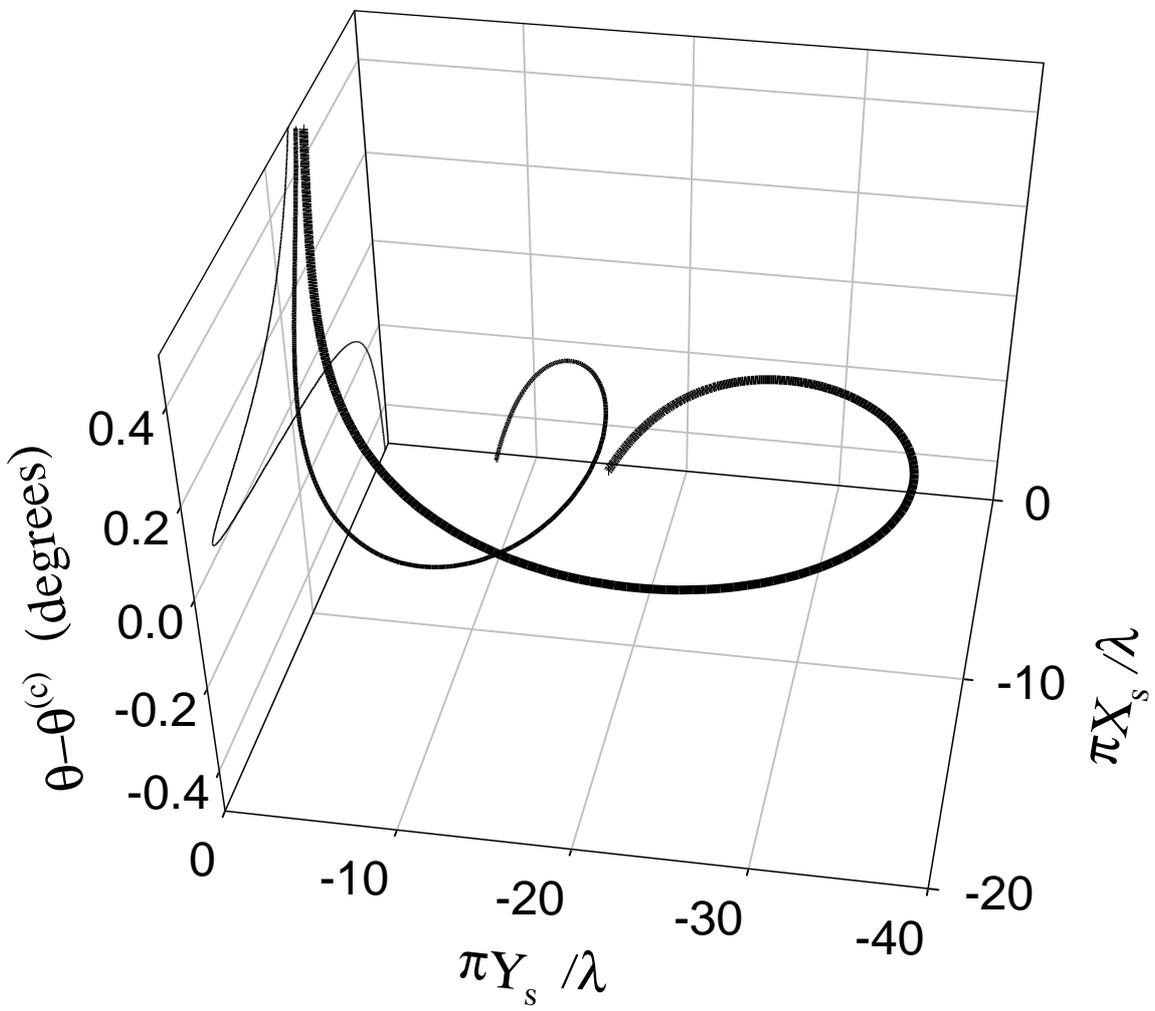
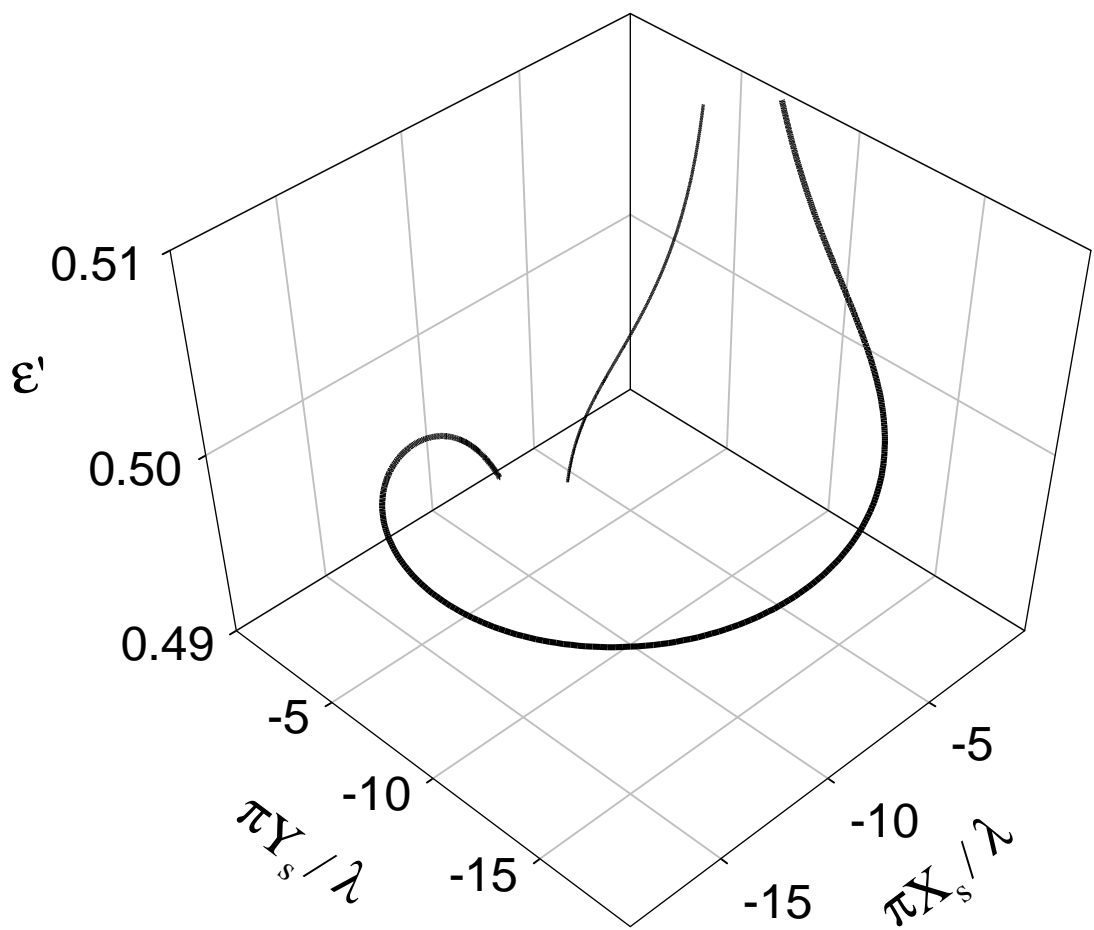
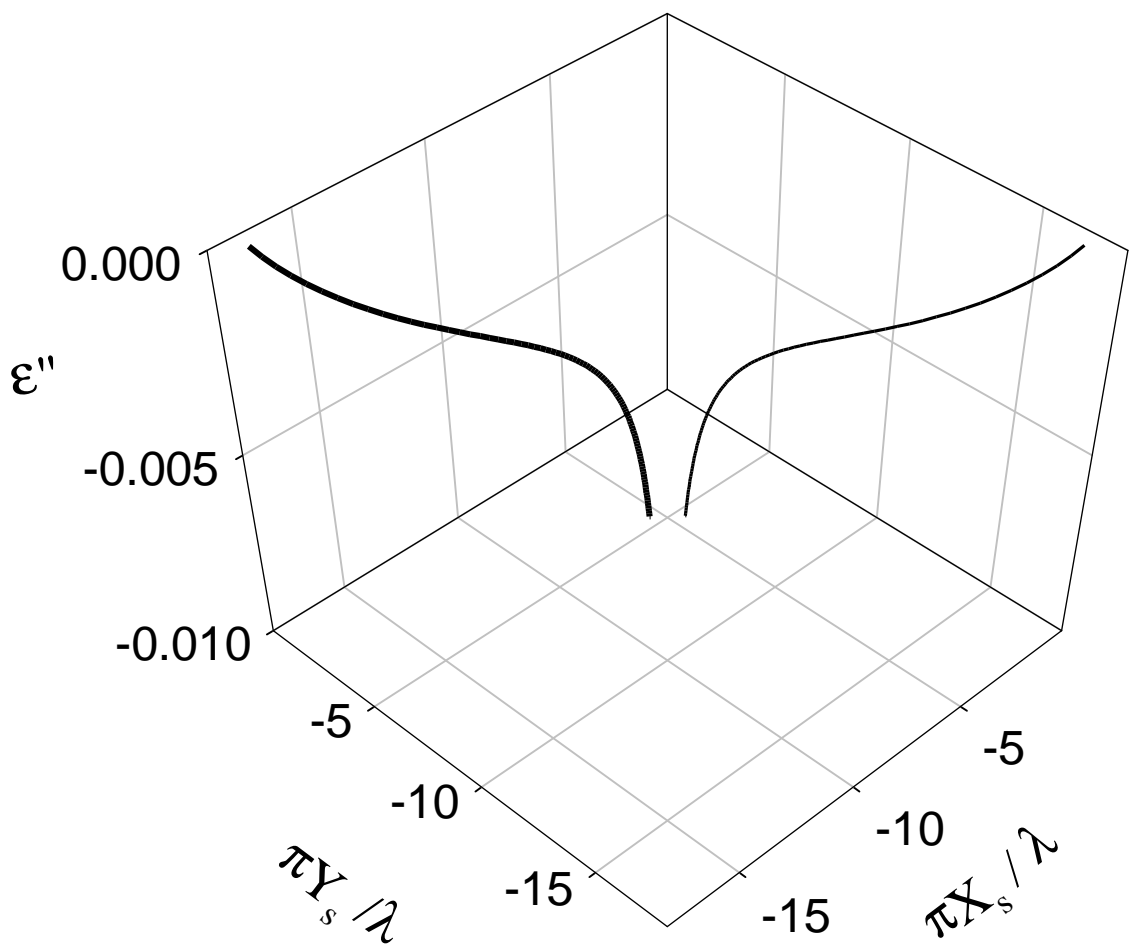
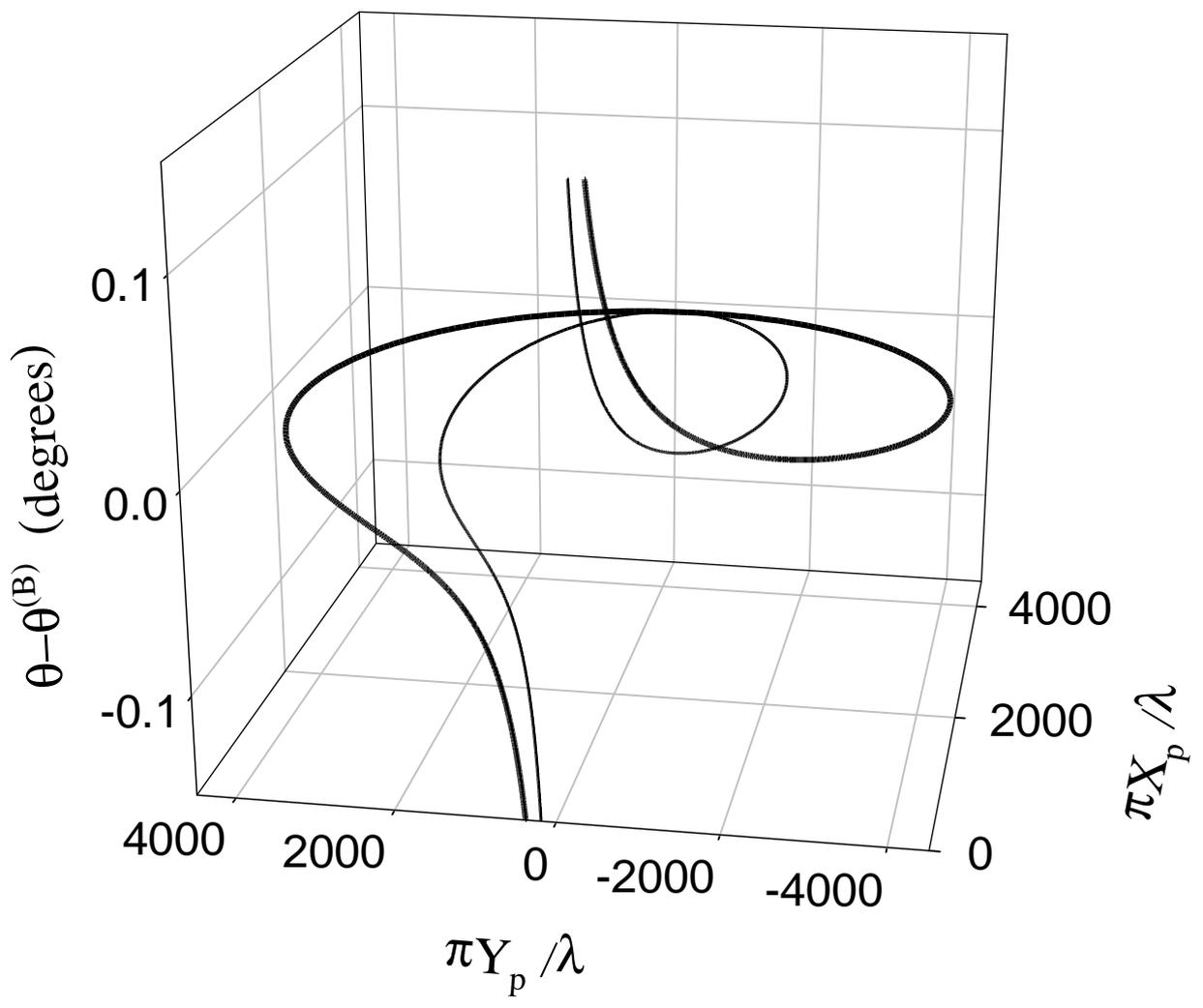
\end{document}